\pdfoutput=1
\documentclass[hyper]{JINST}

\usepackage{graphicx}
\usepackage{amsmath}
\usepackage{wrapfig}
\usepackage[utf8]{inputenc}

\title{Adjoint Monte Carlo Simulation of Fusion Product Activation Probe Experiment in ASDEX Upgrade tokamak}

\author{Simppa Äkäslompolo$^a$\thanks{Corresponding
author.}, Georges Bonheure$^b$, Giovanni Tardini$^c$, Taina Kurki-Suonio$^a$ and The ASDEX Upgrade team\\
\llap{$^a$}Aalto University\\
  Finland\\
\llap{$^b$} ERM-KMS\\Brussels, Belgium\thanks{Present address: European Commision, Research Directorate General}\\
\llap{$^c$}Max-Planck-Institut f\"ur Plasmaphysik, Garching, Germany\\
  E-mail: \email{simppa.akaslompolo@alumni.aalto.fi}}

\abstract{The activation probe is a robust tool to measure flux of fusion products from a magnetically confined plasma. A carefully chosen solid sample is exposed to the flux, and the impinging ions transmute the material making it radioactive. Ultra-low level gamma-ray spectroscopy is used post mortem to measure the activity and, thus, the number of fusion products.

\vspace{1ex}

This contribution presents the numerical analysis of the first measurement in the ASDEX Upgrade tokamak, which was also the first experiment to measure a single discharge. The ASCOT suite of codes was used to perform adjoint/reverse Monte Carlo calculations of the fusion products. The analysis facilitates, for the first time, a comparison of numerical and experimental values for absolutely calibrated flux. The results agree to within a factor of about two, which can be considered a quite good result considering the fact that all features of the plasma cannot be accounted in the simulations.

\vspace{1ex}

Also an alternative to the present probe orientation was studied. The results suggest that a better optimized orientation could measure the flux from a significantly larger part of the plasma.}


\begin{document}

\section{Introduction}

The fast particles, such as fusion alphas, ICRH, as well as NBI generated ions, must be confined in magnetically confined fusion plasmas until they have released their energy. 
Therefore it is essential to measure and understand the fast particle behaviour in burning plasmas, such as in ITER. 
Detecting the escaping particles is the first step in learning to confine them.

The fusion product activation probe is a member of the family of probes and detectors that measure the flux of charged fast particles near the first wall of a magnetic confinement fusion device~\cite{EPphysicsInPreparationForBurningPlasmaExperiments,energeticIonBehaviouInMast2015,fastParticleBehaviourInJETusingNuclearTechniques}. 
The fast ions are often measured with scintillation materials, Faraday cups  and semiconductor detectors 
which can measure on-line the flux with high temporal resolution. 
In contrast, the activation probe is based on post-mortem measurement of the irradiation induced  activity of solid material samples. 
The flux of fast particles transmute part of the target material making it radioactive. 
The resulting gamma activity can be measured with ultra-low level gamma-ray spectroscopy. 
The measurement is well calibrated, which makes the probe particularly well suited for validation of fast particle codes and for cross calibration of other detectors in present day tokamaks. 
The activation probe is very robust, which make it uniquely suitable for the harsh environment of a fusion reactor. 
The activation method requires neither optics nor electronics inside the plasma chamber, where the neutron and gamma radiation would inevitably deteriorate them.
The probe is also inherently isotope selective and practically insensitive to the neutron flux. The measurement can be inherently energy selective, when using an activation process with a threshold energy.

The fusion product flux at the midplane manipulator in ASDEX Upgrade (AUG) was measured for the first time~\cite{bonheureEPS2013} using an activation probe~\cite{actProbe}. \emph{This contribution presents numerical analysis of these experiments using the ASCOT code}~\cite{ascot4ref}. 
In this work, the fusion reactivity is calculated using ASCOT and its AFSI Fusion Source Integrator component, and for verification purposes, also with TRANSP~\cite{TRANSP}. The detailed probe geometry was included in the AUG 3D first wall model. We performed an adjoint Monte Carlo simulation, that was \emph{equivalent} to following the collisionless orbits of the fusion products from their birth location in the plasma to the probe.  In the adjoint method, markers are launched backward in time from the probe towards the plasma. Only fusion protons were experimentally measured, but tritons and helium-3 were modelled for comparison.

The main result of this paper is the comparison of measurements and simulations to demonstrate quantitative analysis capability. Indeed, the current analysis is in reasonable quantitative agreement with the measurements. \emph{The presented numerical analysis provides, for the first time, absolutely calibrated flux of fusion products to the probe.} This is an improvement compared to the previous analysis~\cite{actProbe} performed on JET measurements, that only provided relative profiles of the flux as a function of major radius and thus required external calibration.  Furthermore, the analysis was performed for an alternative probe geometry, as suggested by a previous study~\cite{akaslompoloEPS2013}.

\section{The probe, the measured plasma and the numerical methods}

\begin{figure}
 \centering
   \includegraphics[width=0.4\linewidth]{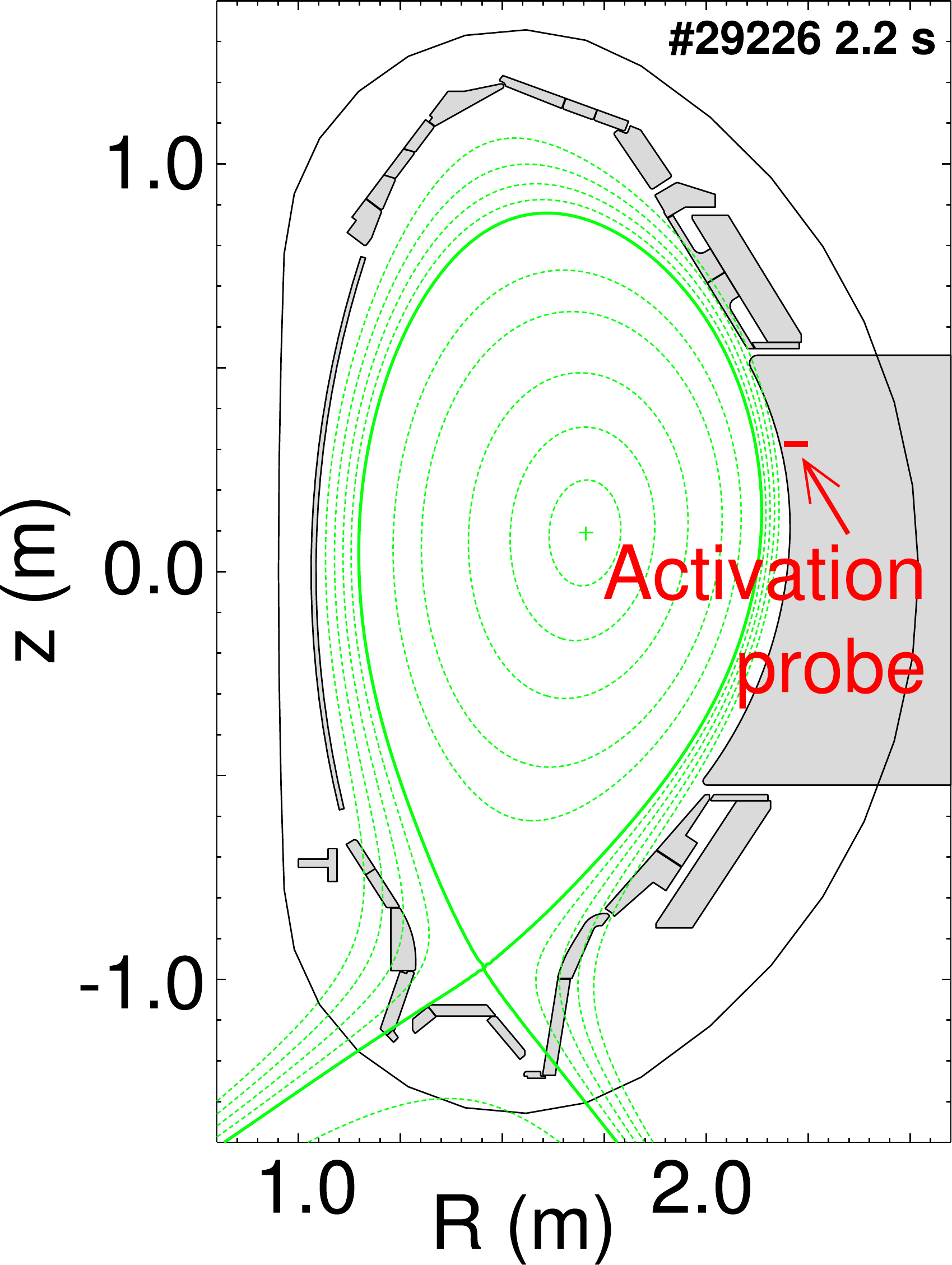}
 \caption{The poloidal cross section of the plasma. The flux surfaces are indicated in green and the activation probe in red.}
 \label{fig:plasmaCrossSection}
\end{figure}

An activation probe exposes carefully chosen material samples to the escaping fusion product flux from the plasma. %
In the AUG xperiments, the probe was connected to the midplane manipulator slightly above the midplane (figure~\ref{fig:plasmaCrossSection}). The geometry is similar to that foreseen for  an  ITER  activation  probe~\cite{actProbeIter}.

\begin{figure}
  \centering
\includegraphics[width=0.75\linewidth]{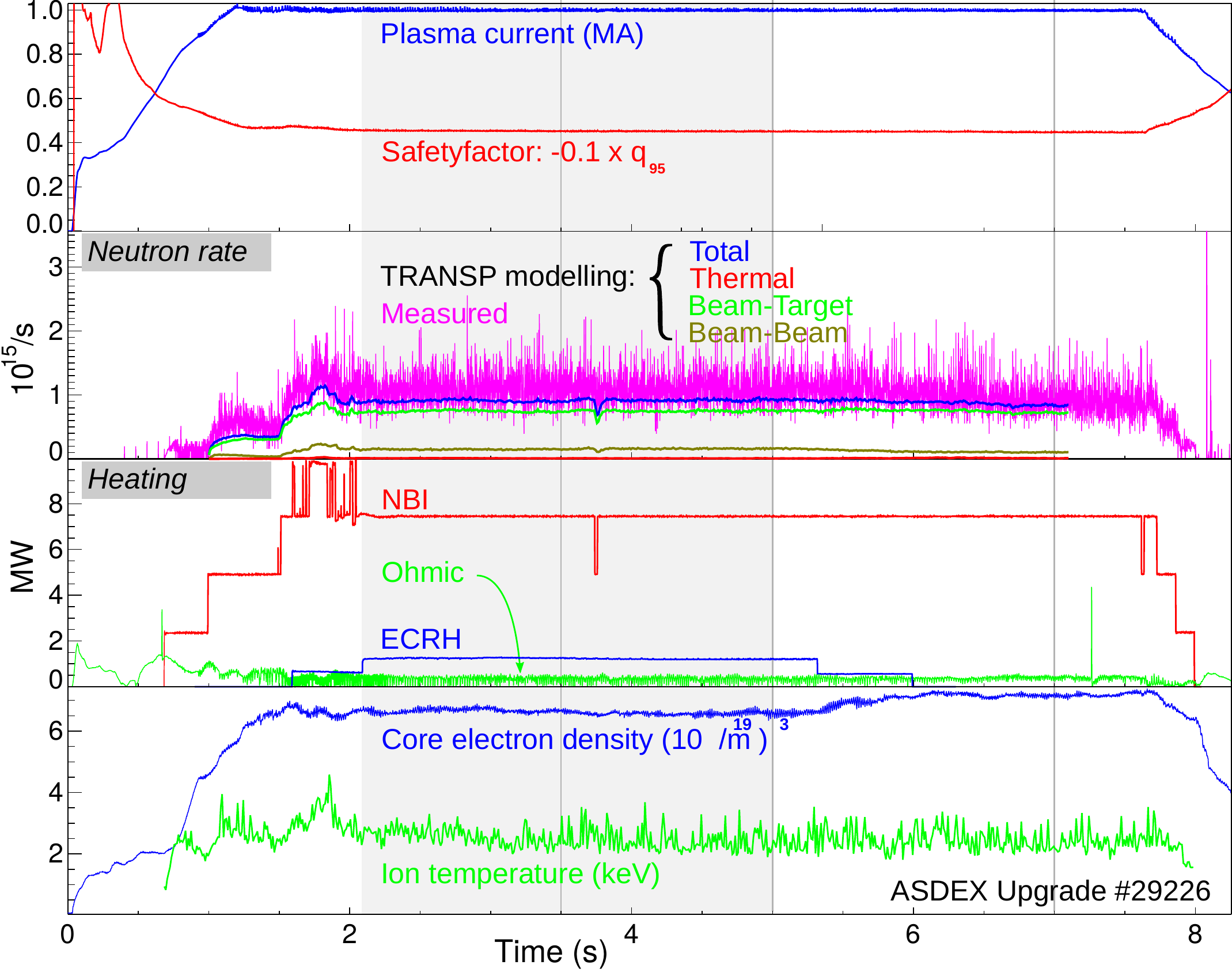}
  \caption{Time evolution of the plasma parameters. The top panel shows the long, uniform plasma flat top. The second panel shows the neutron rate, both measured and one obtained from a transport calculation with TRANSP~\cite{SimulationNeutronRateAUG}. The third panel shows how the neutral beams dominate the heating of the plasma. The bottom panel displays the line averaged electron density from interferometry channel passing through the core plasma and ion temperature from charge exchange recombination spectrometry. The shaded area is the time window for averaging of the background profiles from integrated data analysis~\cite{AUG_IDA}. The gray vertical lines mark the three TRANSP time points, that will be described in section~\protect\ref{sec:results}.  }
  \label{fig:waveform}
\end{figure}

In the experiment, the probe was exposed to the dedicated ASDEX Upgrade discharge \#29226. The plasma flat top was stationary and approximately 6\,s in length (figure~\ref{fig:waveform}). It was a deuterium plasma with 7.25\,MW of injected deuterium neutral beam power. The core temperature was approximately 3\,keV and density $7\cdot10^{19}/\mathrm{m}^{3}$. 
The plasma profiles were acquired using integrated data analysis~\cite{AUG_IDA} for the time window 2.1--5.0\,s, while the magnetic background was calculated for the time point 2.14\,s.  In the modelling, nitrogen was used as the effective impurity, to match the measured effective charge number.

\subsection{The activation probe design and operation principle}
\begin{figure}
  \centering
  (a)\includegraphics[width=0.4\linewidth]{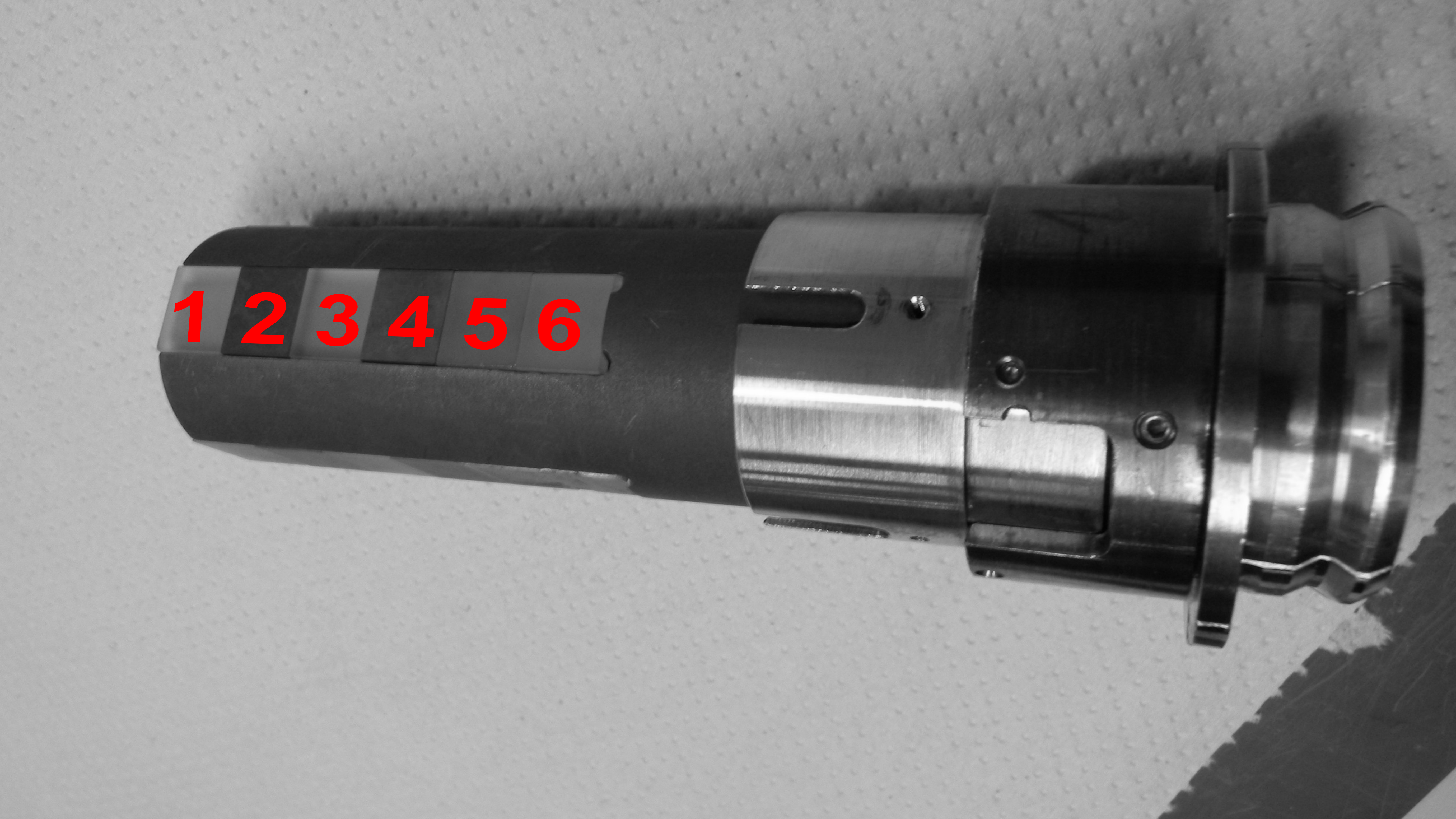} (b)\includegraphics[width=0.4\linewidth]{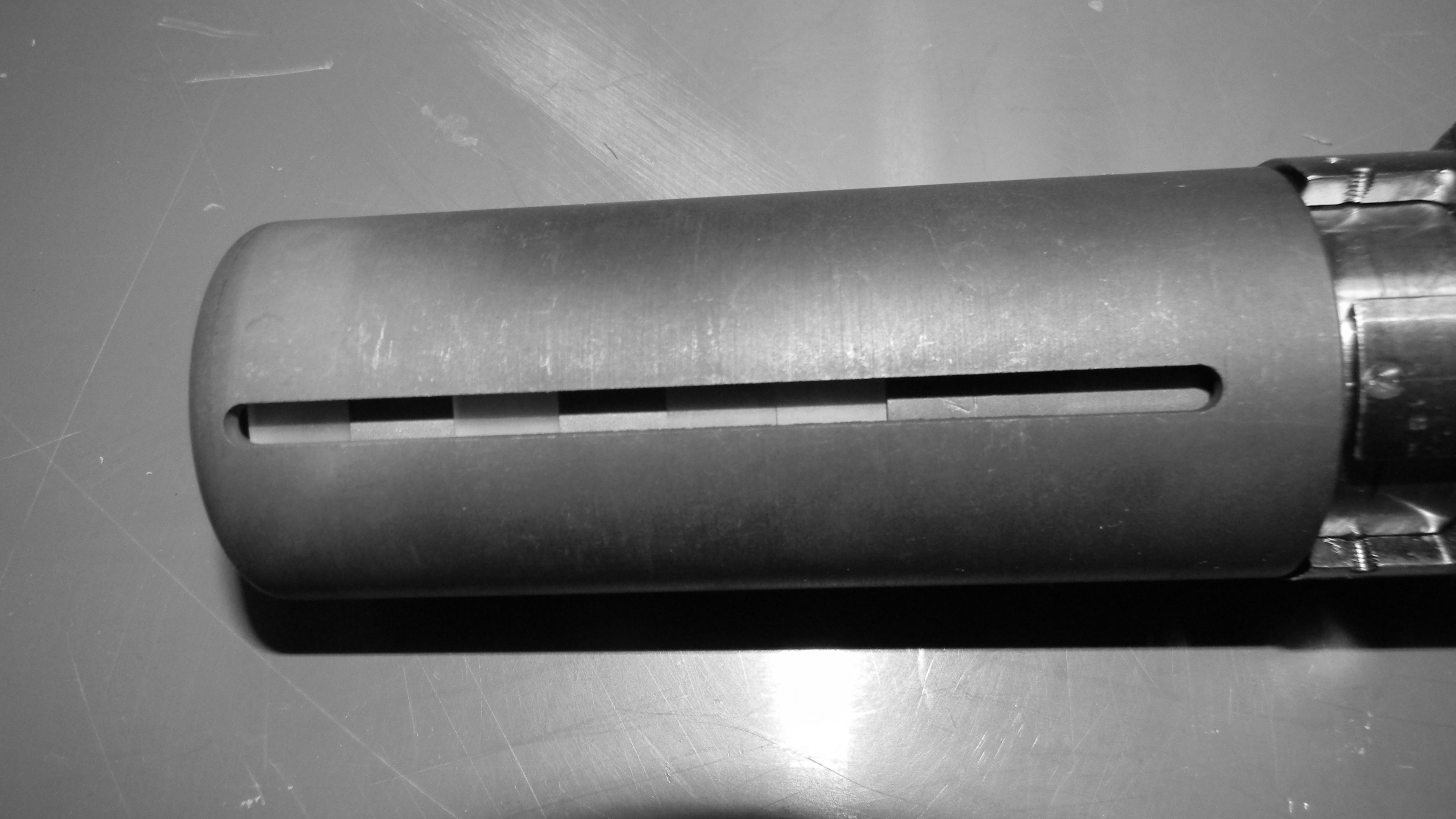}\\
  (c)\includegraphics[width=0.30\linewidth]{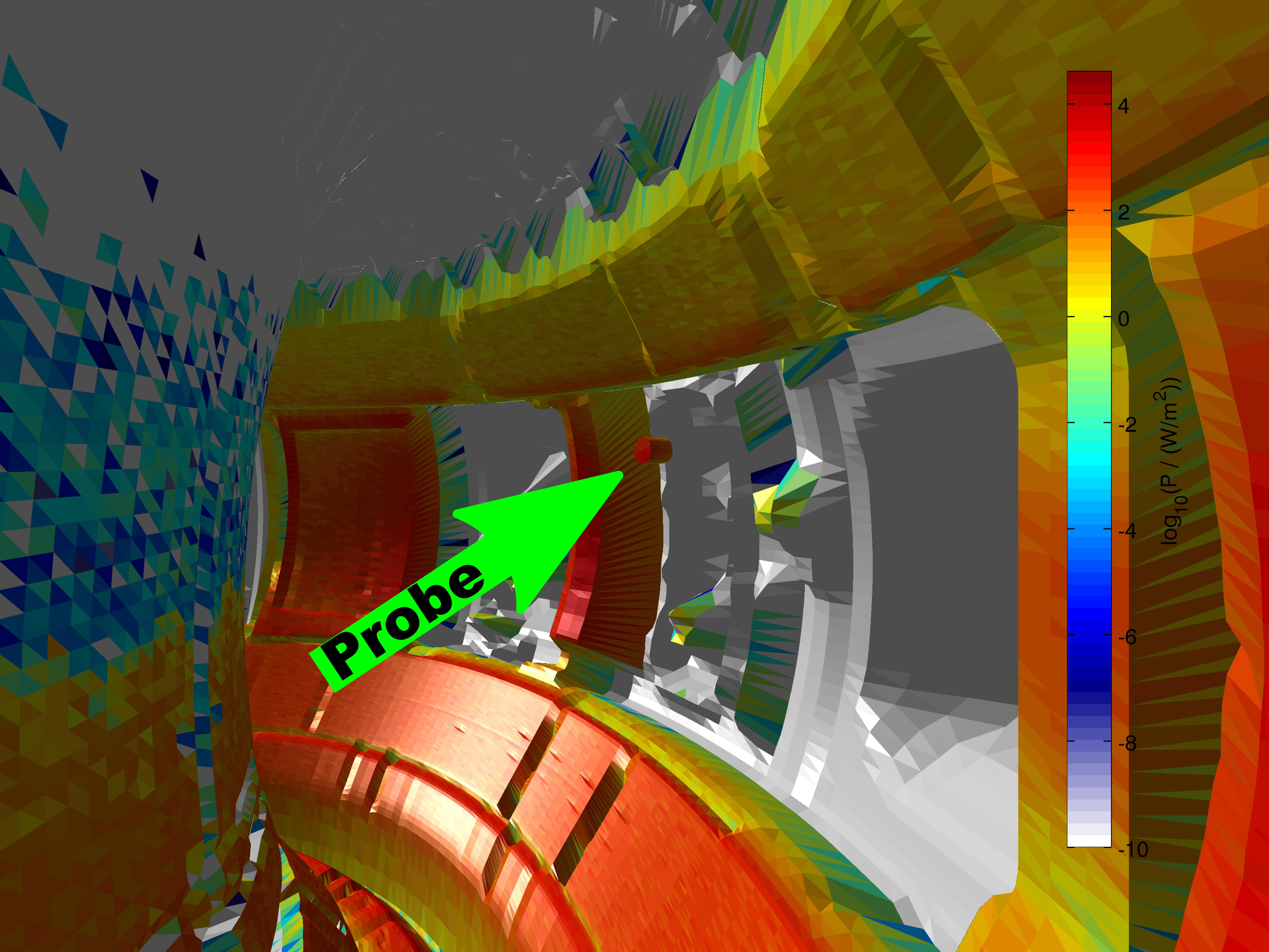} (d)\includegraphics[width=0.275\linewidth,trim=0mm 0mm 0mm 0mm,clip]{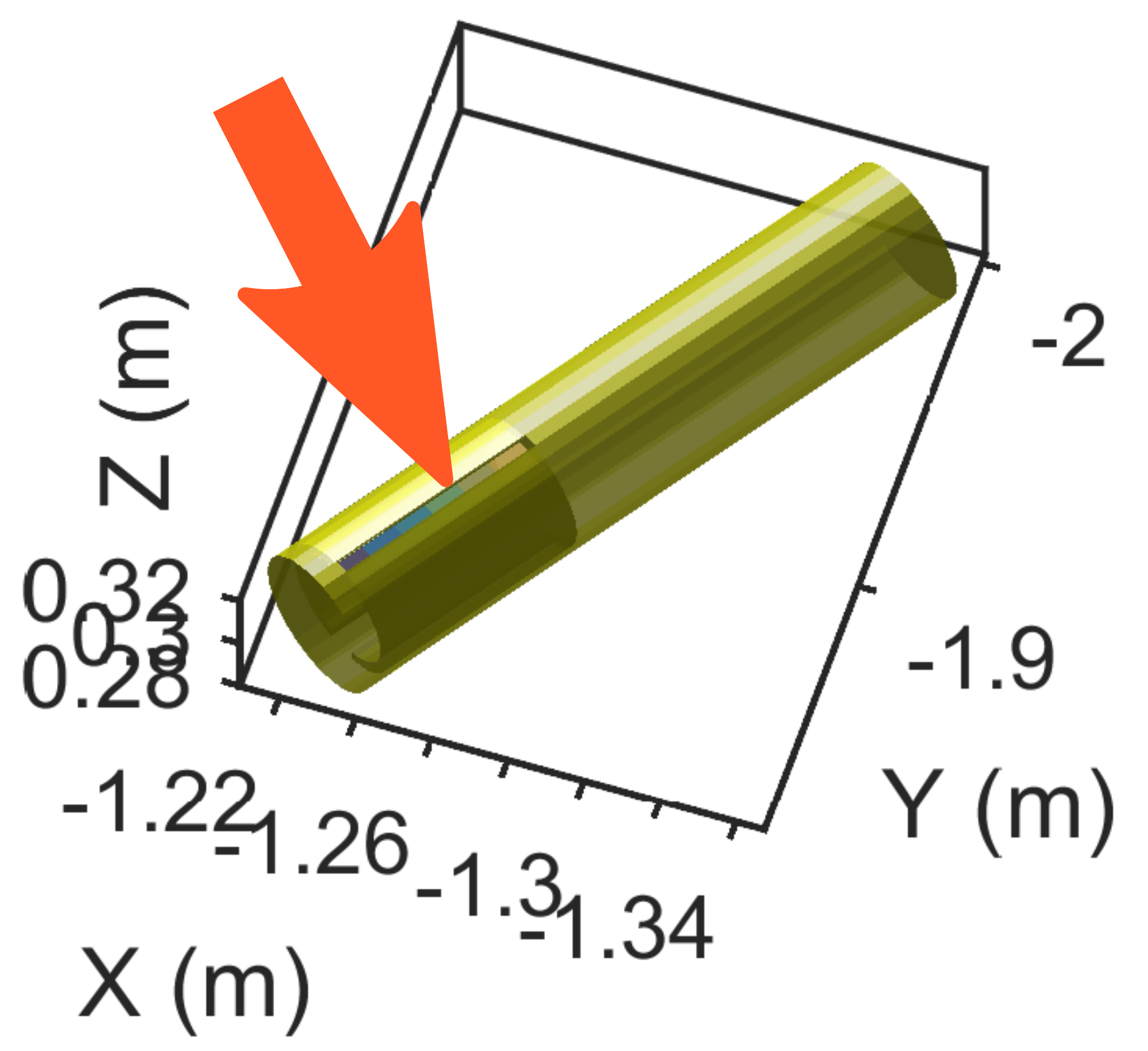} (e)\includegraphics[width=0.30\linewidth,trim=1.5mm 0mm 10mm 0mm,clip]{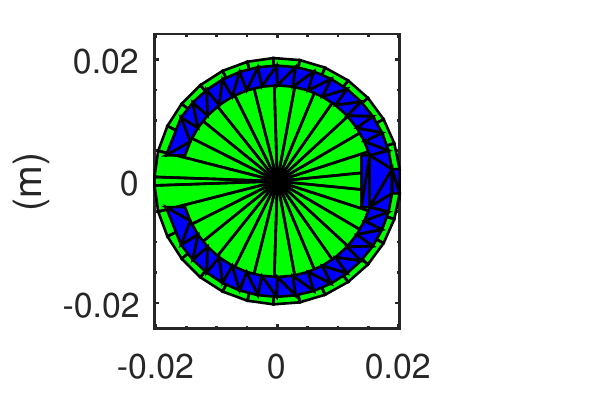}
  \caption{Illustration of the probe. (a) The sample holder with samples numbered. (b) The graphite cap covering the sample holder. (c) The 3D-wall model, including a model for the probe. The coloring indicates fusion product flux distribution onto the first wall as calculated by forward Monte Carlo~\cite{akaslompoloEPS2013}. (d) Close up of the probe geometry used in this study. The protective cover is shown in semi-transparent yellow. The uniquely coloured samples are partly visible through the slit (indicated by the arrow).  (e) Cross-sectional view of the probe. The green mesh is used at the ends of the probe and the blue mesh is used at the ends of the cavity between the sample holder and the graphite cap. The samples and the opening in the cover are at the far right side. }
  \label{fig:probeModel}
  \label{fig:probeIllustrations}
\end{figure}

The probe consists of a graphite sample holder (figure~\ref{fig:probeIllustrations}a) and a protective cover with a slit allowing fusion products to hit the samples (figure~\ref{fig:probeIllustrations}b).  The slit was situated at the top of the graphite cover. Most of the hardware was reused collector probe components~\cite{impuritySurveyCollectorProbe}. %
 Only the sample holder was manufactured for these experiments. 
The holder has four slots, spaced 90$^\circ$ apart, with room for six samples each. The holder can be rotated between discharges to allow measurement of multiple discharges.  The sample materials were in these experiments boron carbide  B$_4$C, calcium fluoride CaF$_2$ and yttrium orthovanadate YVO$_4$.
The two B$_4$C samples (second and fourth, when counting from the tip of the probe) are of interest in this study. The flux of protons to these samples from the D(D,T)p reaction was successfully measured, by measuring the amount of $^7$Be isotope born in the $^{10}$B(p,$\alpha$)$^7$Be nuclear reaction. It is this measurement that will be compared to the simulations. 

In the numerical analysis, the model of the probe was included in the detailed 3D wall (figure~\ref{fig:probeModel}c) of the tokamak. The general shape and dimensions of the graphite cap and sample holder with samples are included in the model as shown in figure~\ref{fig:probeModel}(d-e).

\subsection{The adjoint integration scheme}
The quantity of interest is the flux of particles from the large plasma (source) to the tiny samples within the probe (target). 
The normal, forward in time, calculation is very inefficient, because only very few markers launched from the plasma will find their way to the target that is only visible through the narrow slit in the graphite shell.
This can be remedied by swapping the roles of the source and the target  in the adjoint Monte Carlo integration. 
In this approach, the markers start backward in time from the target and are likely to pass through the plasma before hitting a wall.
In practise, the calculation is done by calculating two independent quantities and subsequently multiplying them to acquire the final flux.
First the fusion product source rate is calculated and then markers are followed backward in time from the samples to the plasma. 
Both calculations are done assuming steady state.

\begin{figure}
  \centering
  \includegraphics[height=7.5cm]{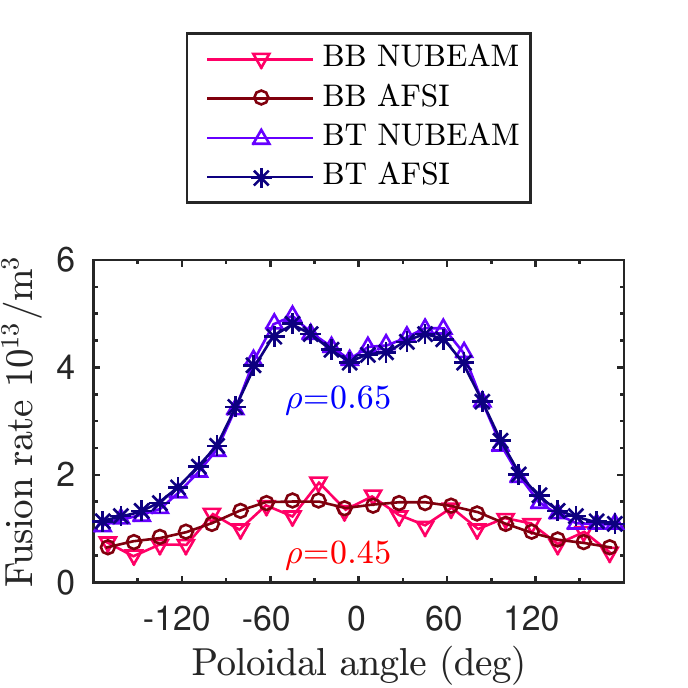}
  \caption{A comparison of fusion reactivity calculated by AFSI and TRANSP/NUBEAM. The plasma profiles and fast ion distributions were identical in this comparison. Beam-beam (BB) reactivity corresponds to reactions that take place among beam ions, while beam-target (BT) reactivity reflects the fusions between thermal plasma and beam ions. Beam target rate is much larger than the BB rate in the studied plasma. The reaction shown here is D(D,n)He$^3$.}
  \label{fig:nubeamAfsi}
\end{figure}

The calculation of fusion reactivity starts with the calculation of the NBI ionisation with the BBNBI code~\cite{Asunta14_BBNBI_reference}. The slowing down distribution of the guiding centers of the beam ions is subsequently calculated with ASCOT~\cite{ascot4ref}, taking into account  collisions with the background plasma. The fusion reactivity is then be calculated using the plasma profiles and slowing down distributions using the AFSI Fusion Source Integrator, a part of the ASCOT suite of codes. The reactivity is stored in a  $(R,z)$ (major radius, elevation) grid. Using identical plasma profiles and slowing down distributions, a very good match (within a few \%), was found between fusion reactivities calculated by AFSI and TRANSP~\cite{TRANSP} (figure~\ref{fig:nubeamAfsi}).

In the second phase of the calculation, the markers are followed backwards in time ($t_\mathrm{prt}$). The initial configuration space marker distribution $A_W$ is uniform on the targets  and the marker velocity distribution $\Omega_W$ is isotropic with a single velocity $v$.
 The full gyro motion of the particle orbits is followed backwards in time until they hit the probe structures or an other part of the tokamak wall. 
\emph{The adjoint density} is the steady-state phase-space density of the reverse-in-time particles as integrated from their trajectories. %
The adjoint density is calculated in a similar grid as the reactivity. 
The flux of fusion products arriving at the target is then obtained by multiplying the two grids element by element and summing them up. This method approximates the following integral for the total flux:
\begin{equation}
  \mathrm{Flux}=\iint\displaylimits_{A_W\Omega_W}\int_{t_\mathrm{prt}}\underbrace{\frac{\left<\sigma u\right>n_1n_2}{4\pi}}_\mathrm{fusion\ reactivity}v\mathrm{d}t\,\mathrm{d}\Omega_W\mathrm{d}A_W
\end{equation}
 For more details, please see~\cite{adjointMC2015_akaslompolo}.

\section{Simulation results and comparison to experiments}

\label{sec:results}

\begin{figure}
  \centering
  \includegraphics[height=5cm]{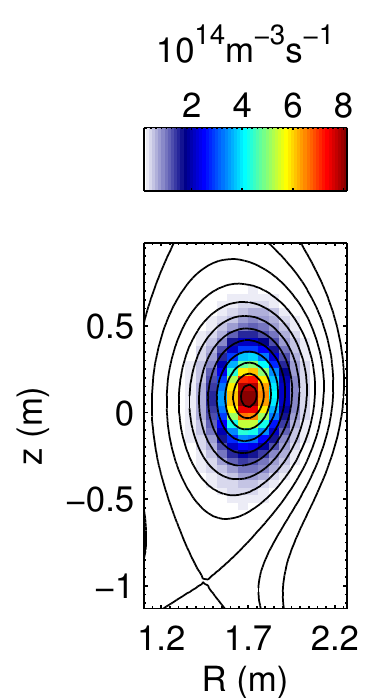}\hspace{-1.9cm}(a)\hspace{2.5cm}\hspace{-1em}
  \includegraphics[height=5cm,trim=13.8mm 0 0 0,clip]{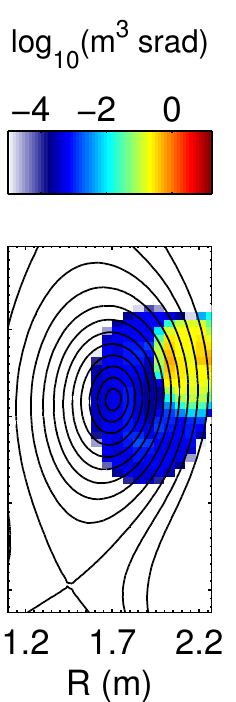}\hspace{-1.9cm}(b)\hspace{1.9cm}\hspace{-1ex}
  \includegraphics[height=5cm,trim=13.8mm 0 0 0,clip]{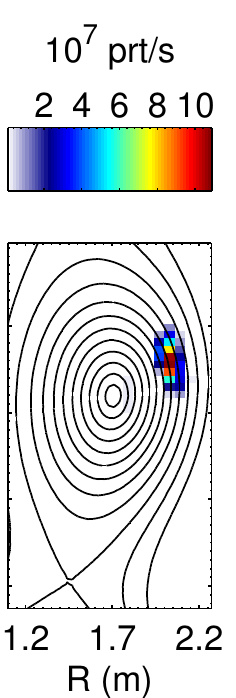}\hspace{-1.9cm}(c)\hspace{1.9cm}\hspace{-1ex}
  \includegraphics[height=5cm,trim=13.8mm 0 0 0,clip]{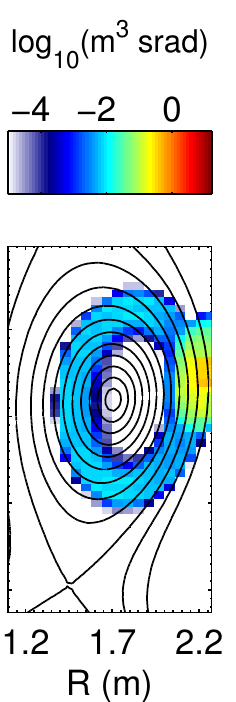}\hspace{-1.9cm}(d)\hspace{1.9cm}\hspace{-1ex}
  \includegraphics[height=5cm,trim=13.8mm 0 0 0,clip]{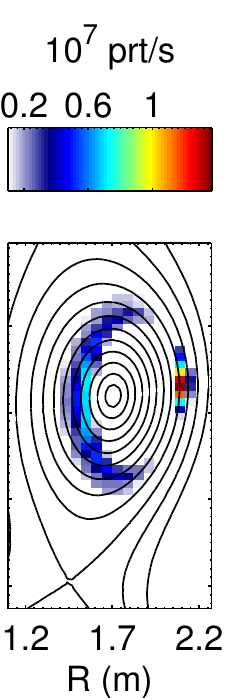}\hspace{-1.9cm}(e)\hspace{1.9cm}\hspace{-1ex}
  \caption{(a) The fusion reactivity. (b) The adjoint marker density of protons, summed up for all samples. (c) The origin of the contributing particles. (d) The adjoint marker density of protons if the probe were rotated by 90$^\circ$. (e)  The birth locations of the contributing particles. (90$^\circ$ rot.)
  }
  \label{fig:resComps}
\end{figure}

The calculation of flux was done in three phases, calculation of the fusion reactivity, calculation of the adjoint density and multiplication of these two together. The fusion reactivity was calculated for thermonuclear, beam-target and beam-beam fusions. The sum of these three reactions is shown in figure~\ref{fig:resComps}(a) for the reaction D(D,p)T. The other reactivity, for the reaction D(D,$^3$He)n, is nearly identical. 

The adjoint calculation was performed by launching three million markers from the samples. Approximately 15\% of the markers exited the graphite cap through the slit. The bulk of these markers hit the wall in less than a microsecond, and produce the high adjoint density in the immediate vicinity of the probe, see figure~\ref{fig:resComps}(b). Even the longest living markers hit a wall within 100 microseconds, during which they produce the volume of low adjoint density that extends to the centre of the plasma. The adjoint density is closely related to the ``instrument function'' of the probe: It directly indicates which parts of the plasma the probe measures and with what kind of relative sensitivity.  In absolute sense, it indicates the ``probing power'' in phase space, which explains the somewhat counter-intuitively units, m$^3\cdot$srad.

\begin{figure}
  \centering
      (a)\includegraphics[width=0.35\linewidth,angle=0]{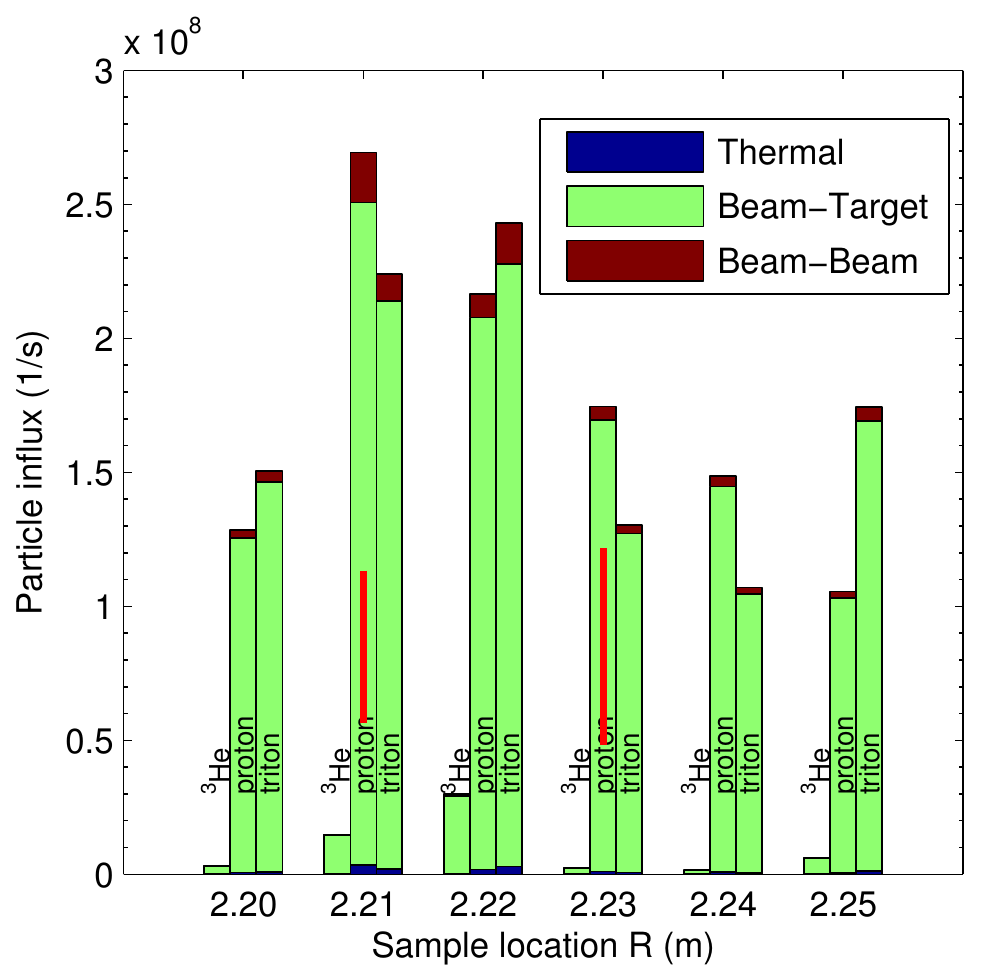} (b)\includegraphics[width=0.36\linewidth]{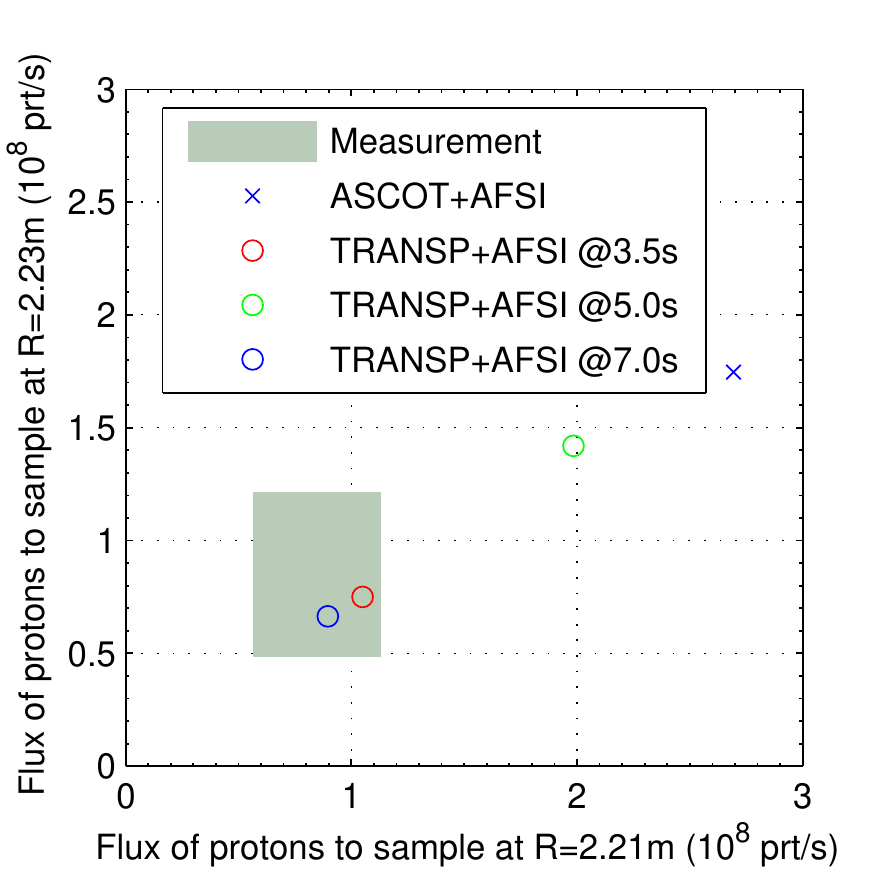}
  \caption{(a) Simulated and measured flux to the samples. In each group of bars, the left one corresponds to $^3$He, the middle one to protons and the right one to tritons. The red bar is the measured proton flux with uncertainty. (b) Comparison of measurements and adjoint simulations using four different fusion reactivities. The green area indicates the measurement uncertainty.  }
  \label{fig:res_top}
  \label{fig:measurements2d}
\end{figure}

The total flux to targets was obtained by multiplying the fusion reactivity by the adjoint density  (figure~\ref{fig:resComps}c), (and scaling with the relevant factors). The fusion reactivity drops quickly outside the core, while the adjoint density is large only near the probe. Consequently their product is non-vanishing mainly at the plasma periphery, at minor radius 0.7<$\rho_\textrm{pol}$<1, at the low field side.  As described above, the resulting flux is sum over the point-wise products and presented in figure~\ref{fig:res_top}(a). The figure also shows the measured flux, obtained by dividing the total number of particles by the flat top length of 6.0\,s. Also a transport simulation of the discharge was performed with TRANSP to verify the forward part of the calculation and to assess the simulation errors. The calculation was repeated for three time steps from the TRANSP simulation (marked in figure~\ref{fig:waveform}). The four calculations are within approximately a factor of two of the experimental measurement as figure~\ref{fig:measurements2d} reveals.

\begin{figure}
  \centering
      (a)\includegraphics[width=0.48\linewidth,trim=5mm 5.5mm 10mm 5mm, clip]{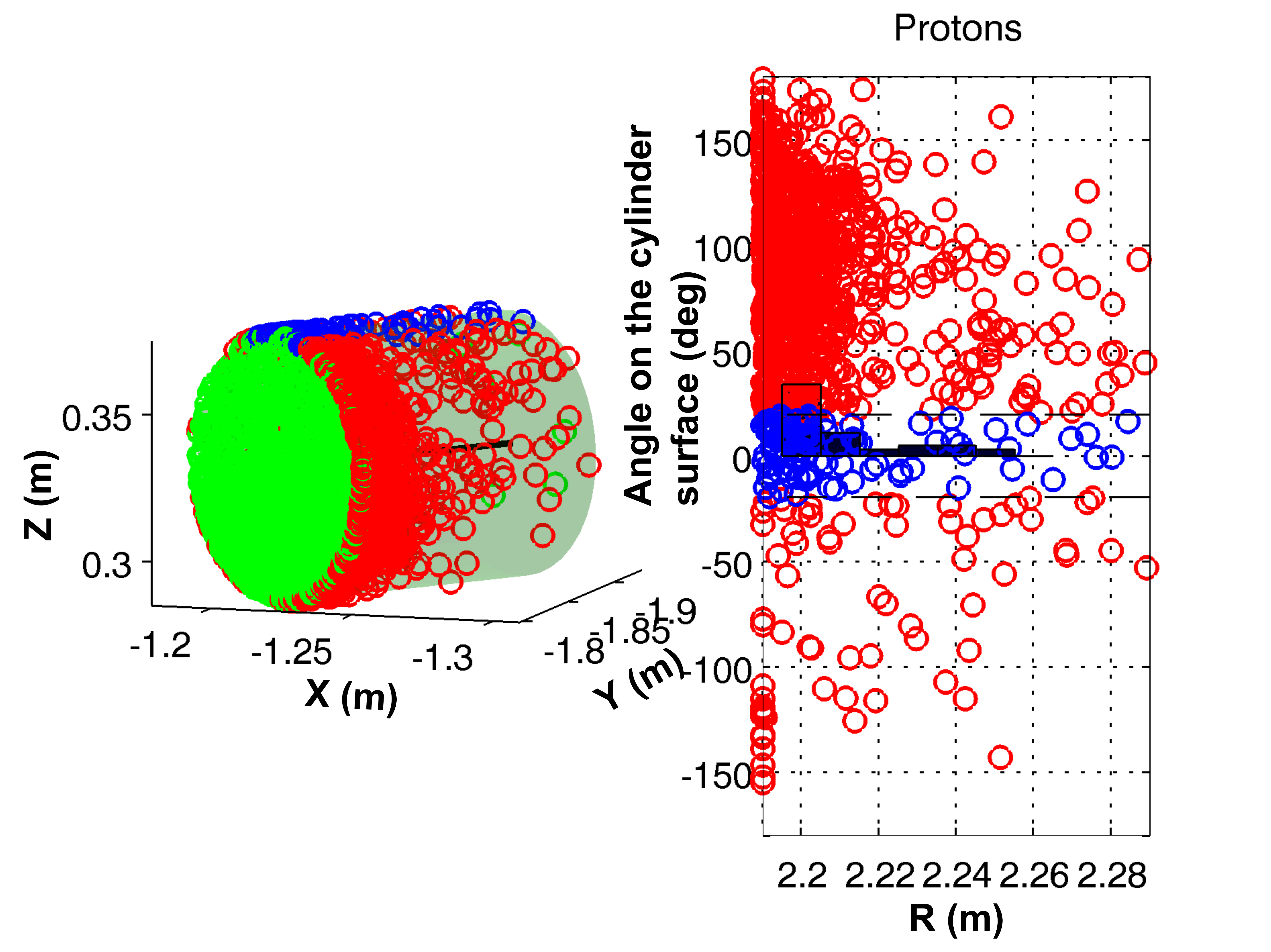} (b)\includegraphics[width=0.4\linewidth,angle=0]{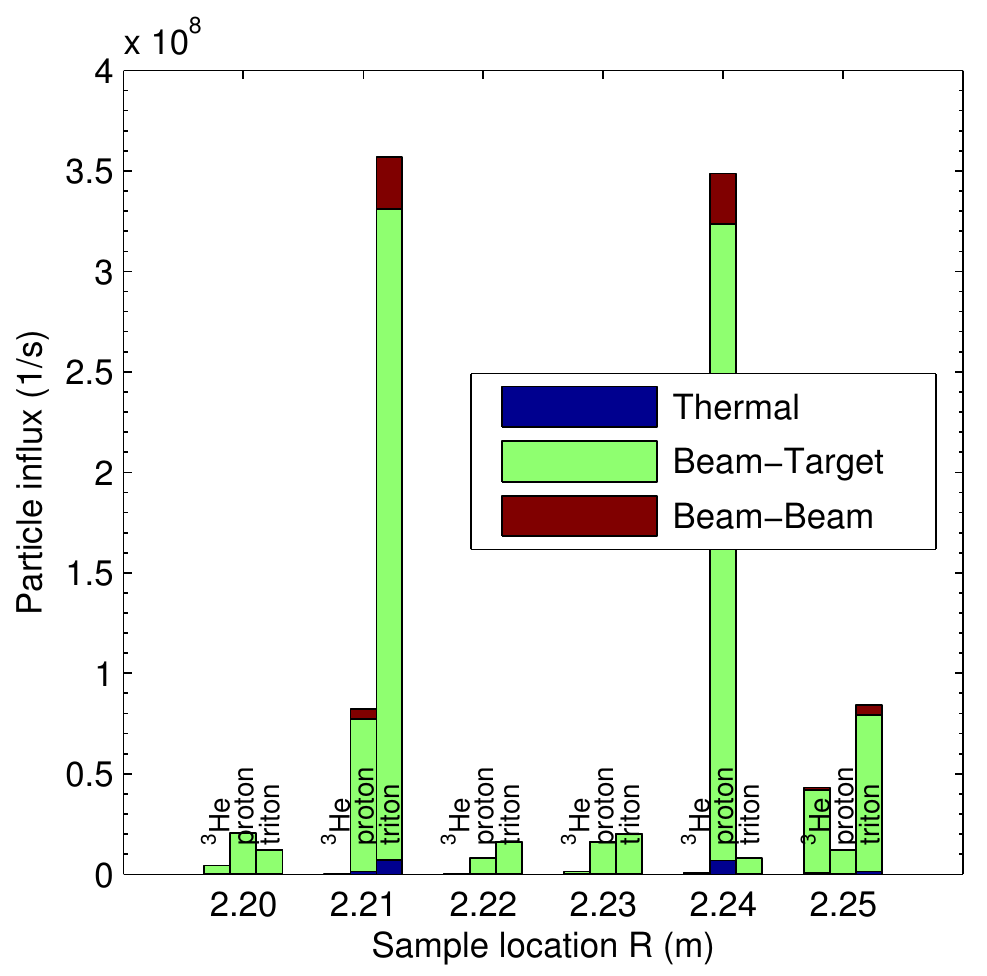}%
  \caption{(a) Forward Monte Carlo simulation results~\cite{akaslompoloEPS2013} indicate that rotating the probe by 90$^\circ$ would have dramatically increased the flux to the samples. The left panel shows model of the probe graphite cover. Each circle depicts a fusion proton hit. The blue ones are approximately the ones that would enter through the slit (at the top of the probe) and green ones have hit the end of the probe. The right panel shows the side of the graphite cover unwrapped onto a plane.  The maximum density of protons is situated approximately 90$^\circ$ from the slit. (b) Adjoint Monte Carlo simulated flux to the samples in case of rotating the probe 90$^\circ$ clockwise as viewed from the plasma, resulting in a horizontal slit. In each group of bars, the left one is $^3$He, the middle one protons and the right one tritons.}
  \label{fig:res_horiz}
\end{figure}

Previous, forward Monte Carlo study~\cite{akaslompoloEPS2013} suggests that if the probe were rotated by 90$^\circ$, the measured fusion product flux would be much higher, because many more products hit the sides of the probe than the top (figure~\ref{fig:res_horiz}a). This was studied with the adjoint method, and indeed, rotating the probe by 90$^\circ$ strongly changed the instrument function (figure~\ref{fig:resComps}d). Rotation of the probe changed the sampled velocity space. In the rotated configuration, the detected products had more parallel velocity when hitting the probe. Such particles were able to evade the walls near the probe and enter from the core of the plasma. Therefore, the probe sampled the fusion reactivity not only at the edge, but also in the core, especially from the high field side 0.3<$\rho_\textrm{pol}$<0.8. The plasma edge contribution from near the probe was reduced, but not completely eliminated (figure~\ref{fig:resComps}e). The combined effect was increase of the signal, but only in certain samples that received the flux from the core (figure~\ref{fig:res_horiz}b). 
 
\section{Summary and Discussion}
We have demonstrated that the activation probe measurements at AUG can be \emph{quantitatively} reproduced using a novel combination of regular Monte Carlo and adjoint Monte Carlo methods, realized by the ASCOT code in this contribution. A successful comparison between the absolutely calibrated measured flux and
the numerically computed value was achieved for the first time. 
The modelling could be improved, by including some of the following features neglected here: the simulations were performed for single time-slice while the plasma evolves in time. The fusion products were assumed to be collisionless. Their birth distribution was assumed to be isotropic and monoenergetic, which is certainly not true for beam-target fusion products. Regardless of these simplifications, the calculated fusion product flux to the probe was within approximately a factor of two of the experimentally measured uncertainty. 

The analysis of rotated probe head suggests that there is large room for optimising the measurement geometry and this contribution describes the tool for the studies. One could conceivably design a probe that is sensitive only to a specific part of plasma. It would be interesting to study the effect of pushing the probe nearer the plasma, rotating the probe, changing the slit shape etc. These studies could also be done for ITER, for which the robustness of the activation probe offers unique possibilities.

\acknowledgments
This work was partially funded by the Academy of Finland project No. 259675. This work  has received funding from Tekes – the Finnish Funding Agency for Innovation under the FinnFusion Consortium.

\bibliographystyle{JHEP.bst}
\bibliography{../bibfile}

\end{document}